\documentclass[aps,prl,showpacs,groupedaddress,superscriptaddress,twocolumn,10pt,nofootinbib,floatfix,preprintnumbers]{revtex4}
\usepackage[dvips]{graphicx}
\unitlength=1.2mm \preprint{JLAB-TH-08-805}
\begin{document}
\newcommand{\tr}{\mbox{tr}\,}
\newcommand{\Dslash}{{\mathchoice
    {\Dslsh \displaystyle}%
    {\Dslsh \textstyle}%
    {\Dslsh \scriptstyle}%
    {\Dslsh \scriptscriptstyle}}}
\newcommand{\Dslsh}[1]{\ooalign{\(\hfill#1/\hfill\)\crcr\(#1D\)}}
\newcommand{\leftvec}[1]{\vect \leftarrow #1 \,}
\newcommand{\rightvec}[1]{\vect \rightarrow #1 \:}
\renewcommand{\vec}[1]{\vect \rightarrow #1 \:}
\newcommand{\vect}[3]{{\mathchoice
    {\vecto \displaystyle \scriptstyle #1 #2 #3}%
    {\vecto \textstyle \scriptstyle #1 #2 #3}%
    {\vecto \scriptstyle \scriptscriptstyle #1 #2 #3}%
    {\vecto \scriptscriptstyle \scriptscriptstyle #1 #2 #3}}}
\newcommand{\vecto}[5]{\!\stackrel{{}_{{}_{#5#2#3}}}{#1#4}\!}
\newcommand{\vdot}{\!\cdot\!}

\bibliographystyle{apsrev}

\title{
\large First Lattice Study of the \boldmath $N$-$P_{11}(1440)$ Transition Form Factors}
\author{Huey-Wen Lin}
\email{hwlin@jlab.org} \affiliation{Thomas Jefferson National
Accelerator Facility, Newport News, VA 23606}
\author{Saul D. Cohen}
\affiliation{Thomas Jefferson National
Accelerator Facility, Newport News, VA 23606}
\author{Robert G. Edwards}
\affiliation{Thomas Jefferson National
Accelerator Facility, Newport News, VA 23606}
\author{David G. Richards}
\affiliation{Thomas Jefferson National
Accelerator Facility, Newport News, VA 23606}

\date{Mar. 19, 2008}

\begin{abstract}
Experiments at Jefferson Laboratory, MIT-Bates, LEGS, Mainz, Bonn, GRAAL, and Spring-8 offer new opportunities to understand in detail how nucleon resonance ($N^*$) properties emerge from the nonperturbative aspects of QCD. Preliminary data from CLAS collaboration, which cover a large range of photon virtuality $Q^2$ show interesting behavior with respect to $Q^2$ dependence: in the region $Q^2 \le 1.5 \mbox{ GeV}^2$, both the transverse amplitude, $A_{1/2}(Q^2)$, and the longitudinal amplitude, $S_{1/2}(Q^2)$, decrease rapidly. In this work, we attempt to use first-principles lattice QCD (for the first time) to provide a model-independent study of the Roper-nucleon transition form factor.
\end{abstract}

\maketitle

{\it Introduction:}
Lattice QCD has successfully provided many experimental quantities from first-principles calculation; however, its success has mostly been restricted to measurements of ground-state quantities. Lattice measurements of excited states could contribute, for example, to hadron spectroscopy, where there are many poorly known states which require theoretical input to be identified. At the EBAC at Jefferson Lab, dynamical reaction models have been developed to interpret extracted $N^*$ parameters in terms of QCD\cite{Lee:2006xu,Matsuyama:2006rp}.

Among these excited nucleon states, the nature of the Roper resonance, $N(1440)$ or $N^\prime$, has been the subject of interest since its discovery in the 1960s. It is quite surprising that the nucleon's excited-state mass is lower than its opposite-parity partner, a phenomenon never observed in meson systems. There are several interpretations of the Roper state, for example, as the hybrid state that couples predominantly to QCD currents with some gluonic contribution\cite{Carlson:1991tg} or as a five-quark (meson-baryon) state\cite{Krehl:1999km}. Some earlier quenched lattice QCD calculations, e.g. Refs.~\cite{Sasaki:2001nf,Guadagnoli:2004wm,Leinweber:2004it,Sasaki:2005ap,Sasaki:2005ug,Burch:2006cc}, find a spectrum inverted with respect to experiment, with $N^\prime$ heavier than the opposite-parity state $S_{11}$. However, in the study of Ref.~\cite{Mathur:2003zf}, where larger lattice box and lighter pion masses are used, their findings of  indicate a rapid crossover of the first positive- and negative-parity excited nucleon states close to the chiral limit. The lattice study has not ruled out that the roper is the first-excited state of nucleon and this is our assumption in this work.

The new data provided from experiments at Jefferson Laboratory, MIT-Bates, LEGS, Mainz, Bonn, GRAAL, and Spring-8 offer a new opportunity to understand in detail how nucleon resonance ($N^*)$ properties emerge from the nonperturbative aspects of QCD. For example, the extraction of the $\gamma N \rightarrow N^*$ transition form factors could help us to understand the dynamical origins of the confinement of constituent quarks and the associated meson cloud. There are various QCD-based hadron models, such as the well developed constituent quark model\cite{Capstick:2000qj,Aznauryan:2007ja} and the covariant model based on Dyson-Schwinger Equations\cite{Maris:2003vk}.

Understanding of the true nature of the Roper may be most easily gained by studying its structure and form factors, such as the nucleon-Roper transition. CLAS collaboration\cite{Mokeev:2006an,Joo:2005gs,Burkert:2004sk,Burkert:2002nr} has studied such transitions induced by electron scattering over a large range of intermediate photon virtuality $Q^2$. In the region $Q^2 \le 1.5 \mbox{ GeV}^2$, both the transverse amplitude, $A_{1/2}(Q^2)$, and the longitudinal amplitude, $S_{1/2}(Q^2)$, drop rapidly in magnitude. This is well described in relativistic quark models with light-cone dynamics, and the sign is consistent with the non-relativistic version. However, in the low-$Q^2$ region, $A_{1/2}(Q^2)$ becomes negative; this is not understood within constituent quark models and requires inclusion of meson degrees of freedom. At the EBAC center, dynamical reaction models have been developed to interpret extracted $N^*$ parameters in terms of QCD\cite{Lee:2006xu,Matsuyama:2006rp}. A model-independent study of these quantities from lattice QCD will serve as valuable help to phenomenologists in analyzing experimental data and will provide better theoretical ground for understanding low-$Q^2$ physics.

{\it The $N(1440) \rightarrow N$ Form Factors: }
From Lorentz symmetry, we expect the matrix element composed of any nucleon states, $N_1$ and $N_2$, to have the following general form:
\begin{eqnarray}
\label{eq:Vector-roper}
\langle N_2\left|V^{\rm }_\mu\right|N_1\rangle_{\mu}(q) &=&
{\overline u}_{N_2}(p^\prime)\left[ F_1(q^2) \left(\gamma_{\mu}-\frac{q_\mu}{q^2}
q\!\!\!/
\right) \right. \nonumber \\
&&\hspace{-1 cm} + \left.
\sigma_{\mu \nu}q_{\nu}
\frac{F_2(q^2)}{M_{N_1}+M_{N_2}}\right]u_{N_1}(p) e^{-iq\cdot x},
\end{eqnarray}
where the equation of motion is used to simplify $-q_\mu \gamma_\mu=M_{N_2}-M_{N_1}$.

To compare with the experimental results, we connect the experimentally measured helicity amplitudes $A_{1/2}$ and $S_{1/2}$ to the transition form factors $F^*_{1,2}$ through
\begin{eqnarray}
\label{eq:helicity}
A_{1/2}(Q^2) &=& k_A(Q^2)G_M(Q^2) \nonumber \\
S_{1/2}(Q^2) &=& k_S(Q^2)G_E(Q^2)
\end{eqnarray}
with
\begin{eqnarray}
k_A(Q^2) &\equiv&
    \sqrt{2\pi \alpha \frac{Q^2+\left(M_{N_1}-M_{N_2}\right)^2}
                           {M_{N_1}\left(M_{N_1}^2-M_{N_2}^2\right)}} \\
k_S(Q^2)  &\equiv&  k_A(Q^2)
          \frac{M_{N_1}+M_{N_2}}{2 \sqrt{2} Q^2 M_{N_2}}
          \sqrt{Q^2+\left(M_{N_1}-M_{N_2}\right)^2} \nonumber \\
 &\times& \sqrt{Q^2+\left(M_{N_1}+M_{N_2}\right)^2},
\end{eqnarray}
the magnetic and electric transition form factors
\begin{eqnarray}
\label{eq:GMGE}
G_M(Q^2) &\equiv& F_1^*(Q^2)+ F_2^*(Q^2) \nonumber \\
G_E(Q^2) &\equiv& F_1^*(Q^2) -\frac{Q^2}{(M_{N_2}+M_{N_1})^2} F_2^*(Q^2),
\end{eqnarray}
and QED running coupling $\alpha \approx 1/137$. Using these definitions, we can reconstruct $F_{1,2}^*$ from experimental values of helicity amplitudes.

{\it Lattice Setup: }
In this exploratory study, we will use an anisotropic lattice; that is, a lattice where the temporal lattice spacing is finer than the spatial ones. It has been demonstrated in the past that for certain calculations, such as glueballs\cite{Morningstar:1999rf} and multiple excited-state masses\cite{Basak:2006ww}, there are great advantages to using anisotropic over isotropic lattices due to the finer lattice time spacing, even when the fundamental constituents are not heavy.

We perform our calculations on quenched\footnote[1]{A ``quenched approximation'' means the effective sea-quark mass is infinite. The quenched QCD (QQCD) is confining, asymptotically free, shows spontaneous chiral symmetry breaking and differs from full QCD only in the relative weighting of the gauge configurations, it is reasonable to use quenched simulations to test new lattice techniques. QQCD is very useful for understanding and controlling sources of errors (except for that due to the quenched approximation) for a new methodology before it is extended to a much more expensive full-QCD calculation.} $16^3 \times 64$ lattices with anisotropy $\xi=3$ (i.e. $a_s=3a_t$), using Wilson gauge action with $\beta=6.1$ and stout-link smeared\cite{Morningstar:2003gk} Sheikholeslami-Wohlert (SW) fermions\cite{Sheikholeslami:1985ij} (with smearing parameters $\{\rho,n_\rho\}=\{0.22,2\}$). The parameter $\nu$ is nonperturbatively tuned using the meson dispersion relation, and the clover coefficients are set to their tadpole-improved values. The inverse spatial lattice spacing is about 2~GeV, as determined by the static-quark potential, and the simulated pion mass is about 720~MeV. In total, we use 200 configurations of anisotropic lattices.

In this work, we will use the variational method and simultaneous fitting on two- and three-point Green functions to extract $N$-$P_{11}$ transition form factors. The nucleon two-point correlators measured on the lattice are
\begin{eqnarray}\label{eq:general_2pt}
\Gamma^{(2)}_{AB}(t;\vec{p})= \sum_n \frac{E_n+M_n}{2E_n} Z_{n,A}
Z_{n,B} e^{-E_n(\vec{p})t},
\end{eqnarray}
where $A$ and $B$ index over the different smearing parameters and $n$ indexes over the basis of nucleon energy eigenstates; the states are defined to be normalized as $\langle 0 |(\chi^N)^\dagger|p,s\rangle = Z u_s(\vec p)$ with (with nucleon interpolating field $\chi^N$); the spinors satisfy
\begin{eqnarray}
\sum_{s} u_s(\vec p)\bar u_s(\vec{p})&=& \frac{E(\vec{p})\gamma^t
-i\vec\gamma\cdot \vec{p} + M}{2 E(\vec{p})}. \label{eq:spinor}
\end{eqnarray}
Traditionally, one is only interested in the ground state; thus the smearing parameters are chosen to overlap as little as possible with excited states. We will use various smearings to create correlators having measurable overlap with the Roper excited state.

Similarly, the three-point function is
\begin{eqnarray}\label{eq:general_3pt}
&&\Gamma^{(3),T}_{\mu,AB}(t_i,t,t_f,\vec{p}_i,\vec{p}_f) =
\sum_n \sum_{n^\prime} \frac{Z_{n^\prime,B}(p_f) Z_{n,A}(p_i)}{Z_V}
  \nonumber \\
\,\,\,&& \times
     e^{-(t_f-t)E_n^\prime(\vec{p}_f)}e^{-(t-t_i)E_n(\vec{p}_i)}
 \times \mbox{ME's}
\end{eqnarray}
where $n$ and $n^\prime$ index energy states and
\begin{eqnarray}\label{eq:ME-expressions}
\mbox{ME's} &=& \sum_{s,s^\prime} {4 E_{n^\prime}(\vec{p}_f)E_n(\vec{p}_i)} T_{\alpha\beta}
                   u_{n^\prime}(\vec{p}_f,s^\prime)_\beta \nonumber \\
&&{}\times \langle N_{n^\prime}(\vec{p}_f,s^\prime)\left|V_\mu\right|N_n(\vec{p}_i,s)\rangle
                   \overline{u}_n(\vec{p}_i,s)_\alpha  \nonumber \\
&=& {\rm Tr}\left[T\cdot
           (E_{n^\prime}(\vec{p}_f)\gamma^t -i\vec\gamma\cdot \vec{p}_f +  M_{n^\prime})
 \right.\nonumber \\
&&{}\cdot \mbox{FF's}
    \cdot \left.(E_{n}(\vec{p}_i)\gamma^t
        -i\vec\gamma\cdot \vec{p}_i+ M_{n})\right],
\end{eqnarray}
where the projection used is $T=\frac{1}{4}(1+\gamma_4)(1+i\gamma_5\gamma_3)$, $Z_V$ is the vector-current renormalization constant and FF denotes the form factors. The vector current in Eq.~\ref{eq:general_3pt} is $O(a)$ on-shell improved with the improved coefficient set to tree-level value\footnote[2]{With smeared fermion actions, it has been seen on three-flavor anisotropic lattices with tree-level tadpole-improved coefficients in the fermion action that the nonperturbative coefficient conditions are automatically satisfied\cite{Lin:2007yf}. Similar behavior has also been observed in a quenched study\cite{Hoffmann:2007nm}, where the nonperturbative coefficients or renormalization constants in a smeared fermion action differed from tree-level values by a few percent.}.
$Z_V$ is calculated nonperturbatively from the isovector vector charge $Z_V=1/g_V^{u-d}$.
Note that when one calculates the three-point Green function in full QCD, there are two possible contraction topologies: ``connected'' and ``disconnected'' diagrams, when the vector current vertex appears on a vacuum bubble. In this work, only ``connected'' quantities are included.

{\it Numerical Results: }
We apply a smearing function with gauge-invariant Gaussian form to improve the overlap of fermion operators with states of interest. To obtain matrix elements involving more than ground states, we need to be careful not to over-smear the fermions; a widely smeared fermion will greatly suppress excited-state signals. (This goes against the standard practice used in the nucleon matrix elements where one increases the smearing parameters to suppress excited states.) We use three Gaussian smearing parameters: $\sigma \in \{0.5, 2.5, 4.5\}$; the largest of these has excellent overlap with just the ground state, while the other two include substantial higher-state contributions. For both two-point and three-point correlators, we calculate all 9 possible source-sink smearing combinations.

From the two-point functions, we wish to extract the energies of at least the nucleon and Roper ($E$'s) and the matrix elements of our lattice operators between the vacuum and the baryon states ($Z$'s). We apply the variational method\cite{Luscher:1990ck} to extract the best principal correlators corresponding to pure energy eigenstates from our matrix of correlators. A simple exponential fit to these yields $M_N=1.48(2)$~GeV and $M_R=2.53(8)$~GeV. The $Z$'s are then determined from the eigenvectors of the principal correlators; we select the eigenvectors from the time $t_Z$ that minimizes the discrepancy between the correlators reconstructed from our $Z$'s and $E$'s and the original three-point data. We find the mass of the negative-parity partner $S_{11}$ using the same method: $M_{S_11}=2.40(10)$~GeV. This mass is lower than the Roper mass due to the high quark masses used, a result consistent with previous lattice calculations; it is expected that the masses will cross when lower quark masses are used.

We calculate three-point functions with source and sink locations at 15 and 48 respectively. (This gives a source-sink separation about 1.1~fm.) The final state always has zero momentum for convenience, while the momentum of the initial state varies. We can get the nucleon form factors, $F_{1}^{NN}$ and $F_{2}^{NN}$, from the large Gaussian smearing ($\sigma=4.5$) runs using a ratio approach\cite{Hagler:2003jd}. We use the ground-state term of the fitting form in Eq.~\ref{eq:general_3pt} on smearing parameter $\sigma=4.5$ three-point correlators. The fit range is adjusted so that the fitted results are consistent under small perturbations to the range. We obtain nucleon-nucleon form factors $F_1$ and $F_2$ consistent with those derived from the ratio method. Note that since our pion mass is far heavier than the physical one, we do not expect to see very good agreement with experiment. As the pion mass approaches lighter values, the lattice data will trend toward the experimental values; for more details, see the recent review in Ref.~\cite{Perdrisat:2006hj}.

The ratio method will not extract matrix elements beyond the ground state, but using the $Z$'s and $E$'s derived from our analysis of the two-point functions, we can extract excited matrix elements from the three-point functions by fitting to the form given in Eq.~\ref{eq:general_3pt}. We want to keep terms in Eq.~\ref{eq:general_3pt} having $n$ and $n^\prime$ running from the ground to the first-excited state; thus, there would be four matrix elements in the minimum expansion, which will be free parameters in our fit. We increase the number of three-point correlators, first using just the diagonal correlators where the smearing is the same at the source and sink, and then using all 9 smearing combinations. The nucleon-nucleon matrix elements are verified against the ratio method, and we check that the transition matrix elements are consistent between different sets. In the rest of this work, we show the results from the full 9-correlator simultaneous fits.

On the lattice, we can obtain both the form factors related to the Roper decay $P_{11} \rightarrow \gamma N$, and the one that is related to photoproduction, $\gamma^* N \rightarrow P_{11}$. A Roper at rest has a $P$-wave decay into a pion and a nucleon and an $S$-wave decay into two pions and a nucleon via $S_{11}\pi$. Since we do not wish to consider the complicated case that occurs when two-particle states may be present, we must avoid kinematical situations in which decays can occur. In our simulation, these are suppressed by the high quark mass and discretization of momentum. However, the lowest-momentum Roper ($E_R \approx 2.7$~GeV) can decay into a lowest-momentum pion ($E_\pi \approx 1.0$~GeV) and a nucleon at rest ($M_N \approx 1.5$~GeV). Since we cannot untangle the Roper from the two-particle state here, we drop this data point. All other decays are forbidden.

\begin{figure}
\begin{center}
\includegraphics[width=0.48\textwidth]{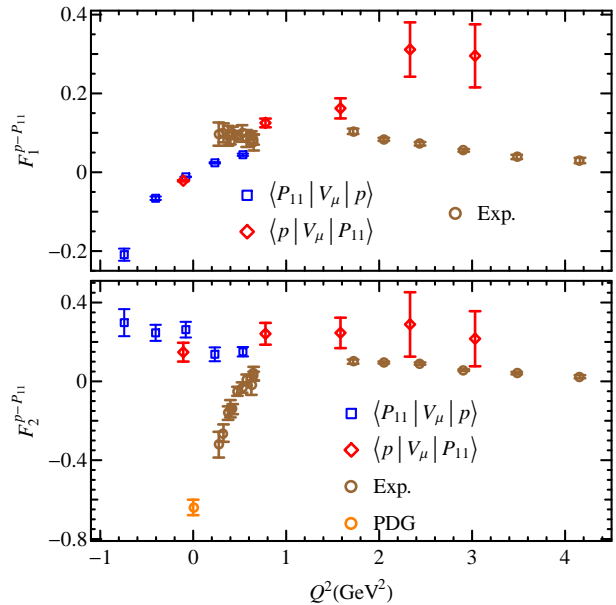}
\end{center}
\caption{\label{fig:F-proton}Proton-Roper form factors $F_{1,2}^*$ obtained from CLAS experiments\cite{Mokeev:2006an,Joo:2005gs,Burkert:2004sk,Burkert:2002nr} and PDG\cite{PDBook} number (circles)  and our fitting method (squares, diamonds)}
\end{figure}

\begin{figure}
\begin{center}
\includegraphics[width=0.48\textwidth]{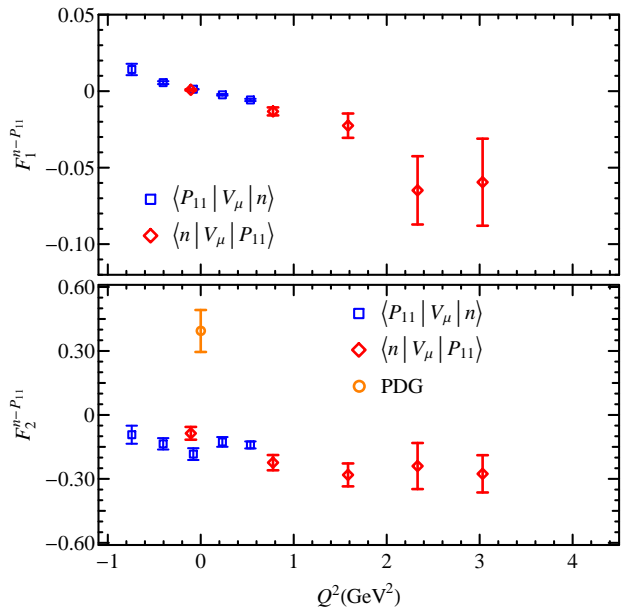}
\end{center}
\caption{\label{fig:F-neutron}Neutron-Roper form factors $F_{1,2}^*$ obtained from PDG\cite{PDBook} number (circles) and our fitting method (squares, diamonds)}
\end{figure}

In Figure~\ref{fig:F-proton}, we show the transition form factors $F_{1,2}^*$ derived from experimental helicity amplitudes\cite{Mokeev:2006an,Joo:2005gs,Burkert:2004sk,Burkert:2002nr} and those from our numbers. In Figure~\ref{fig:F-neutron}, we show the neutron-Roper transition form factors with an experimental point. Note that since our nucleon and Roper masses are much higher than the physical ones, we are in the time-like region when we use the matrix element $\langle P_{11} |V_\mu | N \rangle$ to construct the transition form factors. As we decrease the pion mass, we will enter the space-like region and this matrix element will be helpful in giving us different $Q^2$ points. Our calculation seems to be quite different from the experimental values. This is somewhat expected, since our pion mass is much heavier than the physical value. Even in unquenched calculations, the nucleon form factors do not agree with experimental values with pion mass as low as 300~MeV\cite{Perdrisat:2006hj}. We may expect to see the lattice data approaching experiment as the pion mass used is decreased in the future calculations.

{\it Conclusion and Outlook: }
This exploratory study demonstrates that the nucleon-Roper transition form factors can be measured from a first-principles lattice calculation. Using the fitting approach with appropriately chosen operator smearing, we not only improve the signal in the nucleon-nucleon form factors (especially at large momenta), but also successfully extract the nucleon-Roper. We may vary the projector used in the three-point function to further improve the signal. In the future, since the pion mass in our simulation is very heavy at 720~MeV, we would like to consider lighter pion masses as well as starting work on unquenched anisotropic lattices.

{\it Acknowledgements: }
This work was done using the Chroma software suite\cite{Edwards:2004sx} on clusters at Jefferson Laboratory using time awarded under the SciDAC Initiative. Authored by Jefferson Science Associates, LLC under U.S. DOE Contract No. DE-AC05-06OR23177. The U.S. Government retains a non-exclusive, paid-up, irrevocable, world-wide license to publish or reproduce this manuscript for U.S. Government purposes.



\begin{thebibliography}{29}
\expandafter\ifx\csname natexlab\endcsname\relax\def\natexlab#1{#1}\fi
\expandafter\ifx\csname bibnamefont\endcsname\relax
  \def\bibnamefont#1{#1}\fi
\expandafter\ifx\csname bibfnamefont\endcsname\relax
  \def\bibfnamefont#1{#1}\fi
\expandafter\ifx\csname citenamefont\endcsname\relax
  \def\citenamefont#1{#1}\fi
\expandafter\ifx\csname url\endcsname\relax
  \def\url#1{\texttt{#1}}\fi
\expandafter\ifx\csname urlprefix\endcsname\relax\def\urlprefix{URL }\fi
\providecommand{\bibinfo}[2]{#2}
\providecommand{\eprint}[2][]{\url{#2}}

\bibitem[{\citenamefont{Lee and Smith}(2006)}]{Lee:2006xu}
\bibinfo{author}{\bibfnamefont{T.~S.~H.} \bibnamefont{Lee}} \bibnamefont{and}
  \bibinfo{author}{\bibfnamefont{L.~C.} \bibnamefont{Smith}}
  (\bibinfo{year}{2006}), \eprint{nucl-th/0611034}.

\bibitem[{\citenamefont{Matsuyama et~al.}(2006)\citenamefont{Matsuyama, Sato,
  and Lee}}]{Matsuyama:2006rp}
\bibinfo{author}{\bibfnamefont{A.}~\bibnamefont{Matsuyama}},
  \bibinfo{author}{\bibfnamefont{T.}~\bibnamefont{Sato}}, \bibnamefont{and}
  \bibinfo{author}{\bibfnamefont{T.~S.~H.} \bibnamefont{Lee}}
  (\bibinfo{year}{2006}), \eprint{nucl-th/0608051}.

\bibitem[{\citenamefont{Carlson and Mukhopadhyay}(1991)}]{Carlson:1991tg}
\bibinfo{author}{\bibfnamefont{C.~E.} \bibnamefont{Carlson}} \bibnamefont{and}
  \bibinfo{author}{\bibfnamefont{N.~C.} \bibnamefont{Mukhopadhyay}},
  \bibinfo{journal}{Phys. Rev. Lett.} \textbf{\bibinfo{volume}{67}},
  \bibinfo{pages}{3745} (\bibinfo{year}{1991}).

\bibitem[{\citenamefont{Krehl et~al.}(2000)\citenamefont{Krehl, Hanhart,
  Krewald, and Speth}}]{Krehl:1999km}
\bibinfo{author}{\bibfnamefont{O.}~\bibnamefont{Krehl}},
  \bibinfo{author}{\bibfnamefont{C.}~\bibnamefont{Hanhart}},
  \bibinfo{author}{\bibfnamefont{S.}~\bibnamefont{Krewald}}, \bibnamefont{and}
  \bibinfo{author}{\bibfnamefont{J.}~\bibnamefont{Speth}},
  \bibinfo{journal}{Phys. Rev.} \textbf{\bibinfo{volume}{C62}},
  \bibinfo{pages}{025207} (\bibinfo{year}{2000}), \eprint{nucl-th/9911080}.

\bibitem[{\citenamefont{Sasaki et~al.}(2002)\citenamefont{Sasaki, Blum, and
  Ohta}}]{Sasaki:2001nf}
\bibinfo{author}{\bibfnamefont{S.}~\bibnamefont{Sasaki}},
  \bibinfo{author}{\bibfnamefont{T.}~\bibnamefont{Blum}}, \bibnamefont{and}
  \bibinfo{author}{\bibfnamefont{S.}~\bibnamefont{Ohta}},
  \bibinfo{journal}{Phys. Rev.} \textbf{\bibinfo{volume}{D65}},
  \bibinfo{pages}{074503} (\bibinfo{year}{2002}), \eprint{hep-lat/0102010}.

\bibitem[{\citenamefont{Guadagnoli et~al.}(2004)\citenamefont{Guadagnoli,
  Papinutto, and Simula}}]{Guadagnoli:2004wm}
\bibinfo{author}{\bibfnamefont{D.}~\bibnamefont{Guadagnoli}},
  \bibinfo{author}{\bibfnamefont{M.}~\bibnamefont{Papinutto}},
  \bibnamefont{and} \bibinfo{author}{\bibfnamefont{S.}~\bibnamefont{Simula}},
  \bibinfo{journal}{Phys. Lett.} \textbf{\bibinfo{volume}{B604}},
  \bibinfo{pages}{74} (\bibinfo{year}{2004}), \eprint{hep-lat/0409011}.

\bibitem[{\citenamefont{Leinweber et~al.}(2005)\citenamefont{Leinweber,
  Melnitchouk, Richards, Williams, and Zanotti}}]{Leinweber:2004it}
\bibinfo{author}{\bibfnamefont{D.~B.} \bibnamefont{Leinweber}},
  \bibinfo{author}{\bibfnamefont{W.}~\bibnamefont{Melnitchouk}},
  \bibinfo{author}{\bibfnamefont{D.~G.} \bibnamefont{Richards}},
  \bibinfo{author}{\bibfnamefont{A.~G.} \bibnamefont{Williams}},
  \bibnamefont{and} \bibinfo{author}{\bibfnamefont{J.~M.}
  \bibnamefont{Zanotti}}, \bibinfo{journal}{Lect. Notes Phys.}
  \textbf{\bibinfo{volume}{663}}, \bibinfo{pages}{71} (\bibinfo{year}{2005}),
  \eprint{nucl-th/0406032}.

\bibitem[{\citenamefont{Sasaki et~al.}(2005)\citenamefont{Sasaki, Sasaki, and
  Hatsuda}}]{Sasaki:2005ap}
\bibinfo{author}{\bibfnamefont{K.}~\bibnamefont{Sasaki}},
  \bibinfo{author}{\bibfnamefont{S.}~\bibnamefont{Sasaki}}, \bibnamefont{and}
  \bibinfo{author}{\bibfnamefont{T.}~\bibnamefont{Hatsuda}},
  \bibinfo{journal}{Phys. Lett.} \textbf{\bibinfo{volume}{B623}},
  \bibinfo{pages}{208} (\bibinfo{year}{2005}), \eprint{hep-lat/0504020}.

\bibitem[{\citenamefont{Sasaki and Sasaki}(2005)}]{Sasaki:2005ug}
\bibinfo{author}{\bibfnamefont{K.}~\bibnamefont{Sasaki}} \bibnamefont{and}
  \bibinfo{author}{\bibfnamefont{S.}~\bibnamefont{Sasaki}},
  \bibinfo{journal}{Phys. Rev.} \textbf{\bibinfo{volume}{D72}},
  \bibinfo{pages}{034502} (\bibinfo{year}{2005}), \eprint{hep-lat/0503026}.

\bibitem[{\citenamefont{Burch et~al.}(2006)}]{Burch:2006cc}
\bibinfo{author}{\bibfnamefont{T.}~\bibnamefont{Burch}} \bibnamefont{et~al.},
  \bibinfo{journal}{Phys. Rev.} \textbf{\bibinfo{volume}{D74}},
  \bibinfo{pages}{014504} (\bibinfo{year}{2006}), \eprint{hep-lat/0604019}.

\bibitem[{\citenamefont{Mathur et~al.}(2005)}]{Mathur:2003zf}
\bibinfo{author}{\bibfnamefont{N.}~\bibnamefont{Mathur}} \bibnamefont{et~al.},
  \bibinfo{journal}{Phys. Lett.} \textbf{\bibinfo{volume}{B605}},
  \bibinfo{pages}{137} (\bibinfo{year}{2005}), \eprint{hep-ph/0306199}.

\bibitem[{\citenamefont{Capstick and Roberts}(2000)}]{Capstick:2000qj}
\bibinfo{author}{\bibfnamefont{S.}~\bibnamefont{Capstick}} \bibnamefont{and}
  \bibinfo{author}{\bibfnamefont{W.}~\bibnamefont{Roberts}},
  \bibinfo{journal}{Prog. Part. Nucl. Phys.} \textbf{\bibinfo{volume}{45}},
  \bibinfo{pages}{S241} (\bibinfo{year}{2000}), \eprint{nucl-th/0008028}.

\bibitem[{\citenamefont{Aznauryan}(2007)}]{Aznauryan:2007ja}
\bibinfo{author}{\bibfnamefont{I.~G.} \bibnamefont{Aznauryan}}
  (\bibinfo{year}{2007}), \eprint{nucl-th/0701012}.

\bibitem[{\citenamefont{Maris and Roberts}(2003)}]{Maris:2003vk}
\bibinfo{author}{\bibfnamefont{P.}~\bibnamefont{Maris}} \bibnamefont{and}
  \bibinfo{author}{\bibfnamefont{C.~D.} \bibnamefont{Roberts}},
  \bibinfo{journal}{Int. J. Mod. Phys.} \textbf{\bibinfo{volume}{E12}},
  \bibinfo{pages}{297} (\bibinfo{year}{2003}), \eprint{nucl-th/0301049}.

\bibitem[{\citenamefont{Mokeev and Burkert}(2006)}]{Mokeev:2006an}
\bibinfo{author}{\bibfnamefont{V.~I.} \bibnamefont{Mokeev}} \bibnamefont{and}
  \bibinfo{author}{\bibfnamefont{V.~D.} \bibnamefont{Burkert}},
  \bibinfo{journal}{AIP Conf. Proc.} \textbf{\bibinfo{volume}{842}},
  \bibinfo{pages}{339} (\bibinfo{year}{2006}).

\bibitem[{\citenamefont{Joo et~al.}(2005)}]{Joo:2005gs}
\bibinfo{author}{\bibfnamefont{K.}~\bibnamefont{Joo}} \bibnamefont{et~al.}
  (\bibinfo{collaboration}{CLAS}), \bibinfo{journal}{Phys. Rev.}
  \textbf{\bibinfo{volume}{C72}}, \bibinfo{pages}{058202}
  (\bibinfo{year}{2005}), \eprint{nucl-ex/0504027}.

\bibitem[{\citenamefont{Burkert and Lee}(2004)}]{Burkert:2004sk}
\bibinfo{author}{\bibfnamefont{V.~D.} \bibnamefont{Burkert}} \bibnamefont{and}
  \bibinfo{author}{\bibfnamefont{T.~S.~H.} \bibnamefont{Lee}},
  \bibinfo{journal}{Int. J. Mod. Phys.} \textbf{\bibinfo{volume}{E13}},
  \bibinfo{pages}{1035} (\bibinfo{year}{2004}), \eprint{nucl-ex/0407020}.

\bibitem[{\citenamefont{Burkert}(2003)}]{Burkert:2002nr}
\bibinfo{author}{\bibfnamefont{V.~D.} \bibnamefont{Burkert}},
  \bibinfo{journal}{Eur. Phys. J.} \textbf{\bibinfo{volume}{A17}},
  \bibinfo{pages}{303} (\bibinfo{year}{2003}), \eprint{hep-ph/0210321}.

\bibitem[{\citenamefont{Morningstar and Peardon}(1999)}]{Morningstar:1999rf}
\bibinfo{author}{\bibfnamefont{C.~J.} \bibnamefont{Morningstar}}
  \bibnamefont{and} \bibinfo{author}{\bibfnamefont{M.~J.}
  \bibnamefont{Peardon}}, \bibinfo{journal}{Phys. Rev.}
  \textbf{\bibinfo{volume}{D60}}, \bibinfo{pages}{034509}
  (\bibinfo{year}{1999}), \eprint{hep-lat/9901004}.

\bibitem[{\citenamefont{Basak et~al.}(2006)}]{Basak:2006ww}
\bibinfo{author}{\bibfnamefont{S.}~\bibnamefont{Basak}} \bibnamefont{et~al.}
  (\bibinfo{year}{2006}), \eprint{hep-lat/0609052}.

\bibitem[{\citenamefont{Morningstar and Peardon}(2004)}]{Morningstar:2003gk}
\bibinfo{author}{\bibfnamefont{C.}~\bibnamefont{Morningstar}} \bibnamefont{and}
  \bibinfo{author}{\bibfnamefont{M.~J.} \bibnamefont{Peardon}},
  \bibinfo{journal}{Phys. Rev.} \textbf{\bibinfo{volume}{D69}},
  \bibinfo{pages}{054501} (\bibinfo{year}{2004}), \eprint{hep-lat/0311018}.

\bibitem[{\citenamefont{Sheikholeslami and
  Wohlert}(1985)}]{Sheikholeslami:1985ij}
\bibinfo{author}{\bibfnamefont{B.}~\bibnamefont{Sheikholeslami}}
  \bibnamefont{and} \bibinfo{author}{\bibfnamefont{R.}~\bibnamefont{Wohlert}},
  \bibinfo{journal}{Nucl. Phys.} \textbf{\bibinfo{volume}{B259}},
  \bibinfo{pages}{572} (\bibinfo{year}{1985}).

\bibitem[{\citenamefont{Lin et~al.}(2007)\citenamefont{Lin, Edwards, and
  Joo}}]{Lin:2007yf}
\bibinfo{author}{\bibfnamefont{H.-W.} \bibnamefont{Lin}},
  \bibinfo{author}{\bibfnamefont{R.~G.} \bibnamefont{Edwards}},
  \bibnamefont{and} \bibinfo{author}{\bibfnamefont{B.}~\bibnamefont{Joo}}
  (\bibinfo{year}{2007}), \eprint{arXiv:0709.4680 [hep-lat]}.

\bibitem[{\citenamefont{Hoffmann et~al.}(2007)\citenamefont{Hoffmann,
  Hasenfratz, and Schaefer}}]{Hoffmann:2007nm}
\bibinfo{author}{\bibfnamefont{R.}~\bibnamefont{Hoffmann}},
  \bibinfo{author}{\bibfnamefont{A.}~\bibnamefont{Hasenfratz}},
  \bibnamefont{and} \bibinfo{author}{\bibfnamefont{S.}~\bibnamefont{Schaefer}}
  (\bibinfo{year}{2007}), \eprint{arXiv:0710.0471 [hep-lat]}.

\bibitem[{\citenamefont{Luscher and Wolff}(1990)}]{Luscher:1990ck}
\bibinfo{author}{\bibfnamefont{M.}~\bibnamefont{Luscher}} \bibnamefont{and}
  \bibinfo{author}{\bibfnamefont{U.}~\bibnamefont{Wolff}},
  \bibinfo{journal}{Nucl. Phys.} \textbf{\bibinfo{volume}{B339}},
  \bibinfo{pages}{222} (\bibinfo{year}{1990}).

\bibitem[{\citenamefont{Hagler et~al.}(2003)}]{Hagler:2003jd}
\bibinfo{author}{\bibfnamefont{P.}~\bibnamefont{Hagler}} \bibnamefont{et~al.}
  (\bibinfo{collaboration}{LHPC}), \bibinfo{journal}{Phys. Rev.}
  \textbf{\bibinfo{volume}{D68}}, \bibinfo{pages}{034505}
  (\bibinfo{year}{2003}), \eprint{hep-lat/0304018}.

\bibitem[{\citenamefont{Perdrisat et~al.}(2007)\citenamefont{Perdrisat,
  Punjabi, and Vanderhaeghen}}]{Perdrisat:2006hj}
\bibinfo{author}{\bibfnamefont{C.~F.} \bibnamefont{Perdrisat}},
  \bibinfo{author}{\bibfnamefont{V.}~\bibnamefont{Punjabi}}, \bibnamefont{and}
  \bibinfo{author}{\bibfnamefont{M.}~\bibnamefont{Vanderhaeghen}},
  \bibinfo{journal}{Prog. Part. Nucl. Phys.} \textbf{\bibinfo{volume}{59}},
  \bibinfo{pages}{694} (\bibinfo{year}{2007}), \eprint{hep-ph/0612014}.

\bibitem[{\citenamefont{Yao et~al.}(2006)}]{PDBook}
\bibinfo{author}{\bibfnamefont{W.~M.} \bibnamefont{Yao}} \bibnamefont{et~al.}
  (\bibinfo{collaboration}{Particle Data Group}), \bibinfo{journal}{J. Phys.}
  \textbf{\bibinfo{volume}{G33}}, \bibinfo{pages}{1} (\bibinfo{year}{2006}).

\bibitem[{\citenamefont{Edwards and Joo}(2005)}]{Edwards:2004sx}
\bibinfo{author}{\bibfnamefont{R.~G.} \bibnamefont{Edwards}} \bibnamefont{and}
  \bibinfo{author}{\bibfnamefont{B.}~\bibnamefont{Joo}}
  (\bibinfo{collaboration}{SciDAC}), \bibinfo{journal}{Nucl. Phys. Proc.
  Suppl.} \textbf{\bibinfo{volume}{140}}, \bibinfo{pages}{832}
  (\bibinfo{year}{2005}), \eprint{hep-lat/0409003}.

\end{thebibliography}

\end{document}